\documentclass{aa}
\usepackage{graphicx}

\newcommand{\reff}{\mbox{$r_\mathrm{eff}$}}
\newcommand{\rmax}{\mbox{$r_\mathrm{max}$}}

\newcommand{\hmean}{\mbox{$\langle$h$\rangle$}}

\newcommand{\msun}{\mbox{$M_{\odot}$}}

\newcommand{\ub}{\mbox{$U\!-\!B$}}
\newcommand{\bv}{\mbox{$B\!-\!V$}}
\newcommand{\vi}{\mbox{$V\!-\!I$}}

\begin{document}
\title{Dynamical Mass Estimates for Two Luminous Young Stellar Clusters 
   in \object{Messier 83}
  \thanks{Based on observations collected at the European Southern
          Observatory, Chile under programme 71.B-0303A, and on observations 
	  obtained with the NASA/ESA
    \emph{Hubble Space Telescope}, obtained at the Space Telescope
	  Science Institute, which is operated by the Association
	  of Universities for Research in Astronomy, Inc., under
	  NASA contract NAS 5-26555}
}

\author{S. S. Larsen 
        \inst{1} 
	\and 
	T. Richtler 
	\inst{2}
}

\institute{European Southern Observatory (ESO), ST-ECF,
  Karl-Schwarzschild-Str. 2, D-85748 Garching b. M{\"u}nchen, Germany
  \and
  Universidad de Concepci{\'o}n, Departamento de F{\'i}sica,
  Casilla 160-C, Concepci{\'o}n, Chile
}

\offprints{S.\ S.\ Larsen, \email{slarsen@eso.org}}

\date{Received 29/03/2004; Accepted 23/07/2004}

\abstract{
  Using new data from the UVES spectrograph on the ESO Very Large Telescope and
archive images from the Hubble Space Telescope, we have measured projected 
velocity dispersions and structural parameters for two bright young star 
clusters in the nearby spiral galaxy \object{NGC~5236}. One cluster is located 
near the nuclear starburst of NGC~5236, at a projected distance of
440 pc from the centre, while the other is located in the disk of 
the galaxy at a projected galactocentric distance of 2.3 kpc.
We estimate virial masses for the two clusters of 
$(4.2\pm0.7)\times10^5\msun$ and $(5.2\pm0.8)\times10^5\msun$ and ages (from 
broad-band photometry) of $10^{7.1\pm0.2}$ years and 
$10^{8.0\pm0.1}$ years, respectively. 
Comparing the observed mass-to-light (M/L) ratios with simple stellar 
population models, we find that the data for both clusters are consistent with 
a Kroupa-type stellar mass function (MF).  In particular, we rule out 
any MF with a significantly lower M/L ratio than the Kroupa MF, such as a 
Salpeter-like MF truncated at a mass of 1$\msun$ or higher. These clusters 
provide a good illustration of the fact that massive, globular cluster-like
objects (``super star clusters'') can form at the present epoch even in 
the disks of seemingly normal, undisturbed spiral galaxies.
\keywords{galaxies: star clusters --- galaxies: spiral --- 
  galaxies: individual (NGC~5236)}
}

\titlerunning{Dynamical Mass Estimates for Stellar Clusters in M83}

\maketitle

\section{Introduction}

  Luminous young stellar clusters have been identified in a wide variety of 
external galaxies, including starburst galaxies (van den Bergh \cite{van71}; 
O'Connell et al.\ \cite{oc95}; Meurer et al.\ \cite{meu95}), dwarf irregulars 
(Arp \& Sandage \cite{as85}; Melnick et al.\ \cite{mel85}; 
Billett et al.\ \cite{bhe02}), merger galaxies (Schweizer \cite{schweizer02}; 
Whitmore \cite{whit03}),
nuclear and circumnuclear starbursts (Harris et al.\ \cite{har01};
Maoz et al.\ \cite{maoz01}) and even in the disks of some normal spirals 
(Larsen \& Richtler \cite{lr99}). Many of these clusters are several
magnitudes brighter than any young open cluster 
known in the Milky Way, and the inferred masses can be quite similar to those 
typical of the old globular clusters observed around virtually all large 
galaxies ($10^4-10^6\msun$). This has prompted wide-spread anticipation that 
observations of such clusters may provide clues to the mechanisms that
led to the formation of globular clusters in the early Universe.

  One question which has prompted some debate is whether or not all of these 
young luminous clusters will actually be able to survive for time spans 
comparable to a Hubble time.  While there is ample evidence that at least 
\emph{some} of them do survive for up to several Gyrs, as illustrated e.g.\ 
by the wide range of cluster ages observed in the Large Magellanic Cloud, 
there have been suggestions that some young clusters might have a top-heavy 
stellar mass function (MF). Note that we use the term `MF' rather
than `IMF' to emphasize that we are observing a present-day mass function
which may be different from the initial one.
If a significant fraction of the mass is 
contained in relatively massive stars, the cluster might dissolve once these 
stars reach the endpoint of stellar evolution (Goodwin \cite{goo87}).  
However, the most massive 
young star clusters are quite rare, and therefore tend to be located at 
distances where direct observations of individual low-mass stars are 
impossible with current techniques.  Thus, the MF can only be constrained 
by indirect means. 

  One way to constrain the MF of stellar clusters is to compare a dynamical 
estimate of the cluster mass with theoretical predictions for various MFs, 
based on simple stellar population (SSP) models. Dynamical mass estimates are
usually obtained by measuring the projected velocity dispersion from an 
integrated high-dispersion spectrum of the cluster and combining this with 
information about the physical cluster size (typically derived from an 
HST image), assuming virial equilibrium. This method was first applied
to clusters in NGC~1569 and NGC~1705 (Ho \& Filippenko \cite{hf96a,hf96b}),
and has since been used by several other authors in attempts to constrain the
MF in young clusters. The results so far, however,
are far from clear.  Smith \& Gallagher (\cite{sg01}) found evidence for a
top-heavy MF in the cluster M82-F, while Maraston et al.\ (\cite{mar04})
reported ``excellent agreement'' between the dynamical mass and SSP
model estimates assuming a Salpeter MF extending down to 0.1\msun\ for 
the exceedingly massive cluster W3 in the merger remnant
NGC~7252. For a luminous young cluster in the disk of the nearby spiral 
NGC~6946, Larsen et al.\ (\cite{laretal01}) found a M/L ratio similar to
or even greater than expected for a Salpeter-type MF extending down to
0.1 \msun .  Gilbert \& Graham (\cite{gg03}) found M/L ratios consistent
with ``normal'' MFs for three clusters in NGC~1569.  Other authors have 
found a mixture of mass-to-light ratios 
implying normal MFs in some cases and top-heavy MFs in others 
(Sternberg \cite{s98}, based on a re-analysis of the data by 
Ho \& Filippenko; Mengel et al.\ \cite{men03}; McCrady et al.\ \cite{mgg03}).
Evidence for a top-heavy MF has also been claimed based on Balmer
line strengths measured on medium-resolution spectra for clusters in 
the peculiar galaxy NGC~1275 (Brodie et al.\ \cite{bro98}). 


\begin{figure}
\includegraphics[width=85mm]{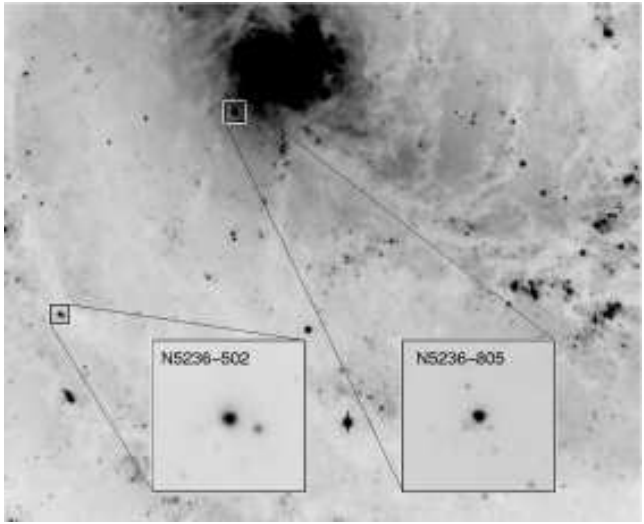}
\caption{A VLT/FORS2 $V$-band image of NGC~5236, showing the locations of 
  the two clusters, and inserts showing HST close-ups. Each insert
  measures $6\arcsec\times6\arcsec$.
}
\label{fig:m83}
\end{figure}

  In this paper we analyse two massive young clusters in the nearby spiral 
galaxy \object{NGC~5236} (M83). In our ground-based survey of young massive 
clusters in nearby spirals (Larsen \& Richtler \cite{lr99,lr00}), NGC~5236 was 
noted as having a very rich population of such clusters, most likely because 
of its high star formation rate.  For consistency with our previous work, 
we refer to the two clusters as N5236-502 and N5236-805.  One of the clusters, 
N5236-805, was already noted in our ground-based data, but the other 
(N5236-502) 
was too red to pass the colour cuts and was noted during later inspection of 
archive images taken with the \emph{Hubble Space Telescope} (HST).
The two clusters both appear well resolved on HST images, so 
that their half-light radii can be reliably measured without being too 
dominated by resolution effects. We adopt a distance modulus for M83 of 
$(m-M)_0 = 28.25\pm0.15$ ($4.5\pm0.3$ Mpc) from Thim et al.\ (\cite{thim03}), 
corresponding 
to a pixel scale on the WF chips of the Wide Field Planetary Camera 2 on 
board HST of 2.2 pc pixel$^{-1}$. Thus, corrections for the WFPC2 point spread 
function (PSF) are still necessary, but should not constitute a major 
uncertainty in the size determinations. The HST images show the
clusters to be reasonably isolated, so we can be confident that the 
velocity dispersions measured from the ground are indeed those of the 
target clusters themselves. For measurements of the velocity dispersions
we use new data from the UVES echelle spectrograph on the ESO Very Large 
Telescope (Sec.~\ref{sec:vdisp}).  To determine the cluster ages we use 
broad-band photometry from 
the Danish 1.54 m telescope at ESO, La Silla, supplemented with archive 
imaging data from the FORS1 and FORS2 instruments on the VLT
(Sec.~\ref{sec:phot}).  The locations 
of the two clusters within NGC~5236 are indicated in Fig.~\ref{fig:m83}.

\section{Velocity dispersions}
\label{sec:vdisp}

\subsection{Data}


\begin{table}
\caption{Template stars. The $h$ values are the peaks of the
  cross-correlation function for the template star spectra vs.\ 
  each of the two clusters, using the range 7310--7570\AA .
  Absolute $M_V$ magnitudes were calculated using apparent
  $V$ magnitudes from the Bright Star Catalogue and 
  \emph{Hipparcos} parallaxes.}
\label{tab:templates}
\begin{tabular}{lcccc} \\ \hline
 Star     &  Sp type & $h$ (502) & $h$ (805) & $M_V$ \\ \hline
HR 4226   &  M1 II   &  0.47     &   0.42    & $-3.4\pm1.1$ \\
HR 4279   &  K1 II   &  0.43     &   0.36    & $-3.2\pm0.8$ \\
HR 5293   &  K4 II   &  0.45     &   0.40    & $\ldots$ \\
HR 5547   &  G8 II   &  0.43     &   0.38    & $-1.2\pm0.3$ \\
HR 5645   &  K4 Ib   &  0.47     &   0.42    & $\ldots$     \\
HR 5738   &  G2 II   &  0.41     &   0.37    & $-2.2\pm0.7$ \\
HR 6120   &  G8 Ib   &  0.46     &   0.41    & $-2.3\pm0.8$ \\
HR 6693   &  M1 Ib   &  0.43     &   0.42    & $\ldots$ \\
HR 7139   &  M4 II   &  0.42     &   0.43    & $-2.9\pm0.3$ \\ \hline
\end{tabular}
\end{table}

\begin{table*}
\caption{Velocity dispersions (km/s). Mean values: (1) Weighted by
  (1/rms), (2) weighted by \hmean . S/N is per resolution
  element (3 pixels). Wavelength ranges: 4520--4850\AA\ (UVES-B),
  6770--6850\AA\ and 6940--7110\AA\ (UVES-RL(1)),
  7310--7570\AA\ (UVES-RL(2)),
  7710--8070\AA\ (UVES-RL(3)), and
  8680--8880\AA\ (UVES-RU).
  }
\label{tab:vd}
\begin{tabular}{lcccccccccc} \\ \hline
             & \multicolumn{5}{c}{N5236-502}         & \multicolumn{5}{c}{N5236-805} \\
	     & Mean & Median & r.m.s & \hmean & S/N  & Mean & Median & r.m.s & \hmean & S/N  \\ 
	     & km/s & km/s   & km/s &        &      & km/s &  km/s  & km/s &        &      \\ \hline
UVES-B       & 5.72 & 5.77   & 0.86 & 0.17  &  15   & 8.46 & 8.46 & 0.47 & 0.12 & 27 \\ 
UVES-RL(1)   & 5.23 & 5.15   & 0.54 & 0.28  &  35   & 9.57 & 9.52 & 0.87 & 0.17 & 36 \\
UVES-RL(2)   & 5.51 & 5.42   & 0.40 & 0.43  &  32   & 7.82 & 7.81 & 0.34 & 0.39 & 34 \\
UVES-RL(1,2) & 5.36 & 5.34   & 0.39 & 0.35  &   -   & 8.22 & 8.26 & 0.31 & 0.28 &  - \\
UVES-RL(3)   & 4.90 & 5.04   & 0.61 & 0.28  &  27   & 7.67 & 7.69 & 0.59 & 0.20 & 29 \\
UVES-RU      & 6.33 & 6.28   & 0.81 & 0.34  &  24   & 7.90 & 7.98 & 0.82 & 0.42 & 28 \\
Mean$^1$     & $5.42\pm0.15$ & $5.39\pm0.13$ & & &
              & $8.13\pm0.18$ & $8.14\pm0.18$ & & &        \\ 
Mean$^2$     & $5.53\pm0.20$ & $5.50\pm0.18$ & & &
              & $8.02\pm0.19$ & $8.05\pm0.18$ & & &        \\ 
Adopted      & \multicolumn{2}{l}{$5.5\pm0.2$} & & & &
                \multicolumn{2}{l}{$8.1\pm0.2$} & & \\ \hline
\end{tabular}
\end{table*}

\begin{figure}
\includegraphics[width=85mm]{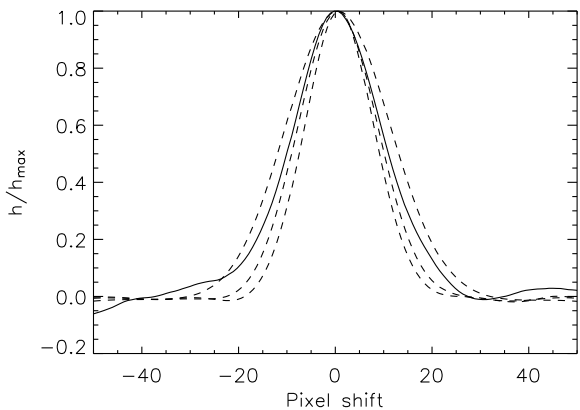}
\caption{Illustration of the cross-correlation technique. The solid curve
  shows the cross-correlation function (CCF) for cluster N5236-502 vs.\
  the template star HR 4226. The three dashed curves show the CCFs for
  template star HR 5645 broadened by
  0 km/s, 4 km/s and 8 km/s, vs.\ HR 4226. For clarity, all CCFs have
  been shifted to a mean of 0 and normalised to a peak value of 1.
}
\label{fig:pccf}
\end{figure}

  For the velocity dispersion measurements we obtained spectroscopic 
observations in service mode with the UVES 
echelle spectrograph on the VLT / UT2 (Kueyen) at ESO, Paranal, in 
March 2003.  We used a slit width of $0\farcs6$, providing a spectral 
resolution $\lambda/\Delta\lambda\sim60,000$, and a dichroic beam splitter
to obtain simultaneous exposures with the blue side (3730\AA -- 4990\AA)
and red side (6600\AA -- 10600\AA ) of UVES.  The UVES blue side is covered 
by a single CCD (UVES-B), while the red side uses a mosaic of two CCDs 
(UVES-RL and UVES-RU), resulting in a small gap in the wavelength coverage.  
In addition to the two clusters in NGC~5236, one cluster in NGC~2997 was
observed, but this spectrum turned out to have 
too low S/N for our analysis.  We also observed a number of late-type 
bright giants and supergiant template stars, listed in 
Table~\ref{tab:templates}. 
Estimates of the absolute $M_V$ magnitudes are listed for each star,
based on the apparent $V$ magnitudes in the Bright Star Catalogue
(Hoffleit \& Jaschek \cite{hoff91}) and \emph{Hipparcos} parallaxes
(Perryman et al.\ \cite{per97}), assuming zero reddening towards these
stars. While a parallax is available for only one luminosity class Ib 
star (HR 6120), it is curious that this star appears to be no more luminous 
than the luminosity class II stars in the table.  Part of the reason may be 
that the relative uncertainties on the parallaxes are substantial, as
reflected in the large uncertainties on the $M_V$ values. Another 
factor may be our neglect of reddening. The Bright Star Catalogue lists
a \bv\ colour of \bv = 1.48 for HR 6120, while G8 supergiants have
an intrinsic $(\bv)_0 \approx 1.2$ (e.g.\ Schmidt-Kaler \cite{sk82}). This
suggests a foreground reddening of $E(\bv)\approx0.3$ or $A_V\approx1$ mag.
The Galactic extinction maps of Schlegel et al.\ (\cite{sch98}) indicate
an extinction of $E(B-V)=0.452$ mag in the direction of HR 6120, which
may be considered an upper limit to the extinction between us and the
star itself but is again consistent with $A_V\approx1$ mag towards HR 6120.

The total exposure time for N5236-502 was 11680 s (195 min), 
divided into 4 separate integrations. For N5236-805, the exposure time was 
7500 s (125 min), divided into 3 integrations.  The initial reductions were 
done with the standard UVES pipeline which bias-subtracts and flatfields
the data, extracts the spectra, performs 
wavelength- and flux calibration and merges the echelle orders to a single 
spectrum for each CCD.  The final reduced cluster spectra were 
co-added, using a sigma-clipping algorithm to remove any remaining cosmic 
ray hits. Typical S/N values for the co-added spectra were $\sim$ 25--35 per 
resolution element (Table~\ref{tab:vd}).  

\subsection{Analysis}

  To measure the velocity dispersions of the clusters we
employed the cross-correlation technique described by Tonry \& Davis 
(\cite{td79}). This method has previously been used in several other
studies of young clusters, including Ho \& Filippenko (\cite{hf96a,hf96b})
and Larsen et al.\ (\cite{laretal01}).  For the cross-correlation we
used the FXCOR task in the RV package in IRAF\footnote{IRAF is distributed
by the National Optical Astronomical Observatories, which are operated by
the Association of Universities for Research in Astronomy, Inc.~under
contract with the National Science Foundation}.  Although the UVES pipeline 
produces flux-calibrated spectra, the signature of the individual echelle 
orders was still clearly visible. To prevent these variations from producing 
spurious features in the cross-correlation function (CCF), we fitted a 
high-order spline to the continuum level and filtered out low spatial 
frequencies with a ramp filter before the cross-correlation analysis.  


  Figure~\ref{fig:pccf} illustrates the basic idea behind our
implementation of the cross-correlation 
method: first, the cluster spectrum is cross-correlated with the spectrum of a 
suitable template star. The full-width-at-half-maximum (FWHM) of the peak of 
the resulting CCF is related to the broadening of the spectral lines,
here assumed to be due to the motions of the stars
within the cluster. The relation between the FWHM of the CCF peak and 
the velocity dispersion is established by broadening another template star
spectrum with Gaussians corresponding to different velocity dispersions,
and cross-correlating the broadened spectra with the first template. This
method assumes that the projected velocity distributions are indeed 
Gaussian, which is expected to be a good approximation for relaxed clusters 
with isotropic velocity distributions (King \cite{king65}).
In Fig.~\ref{fig:pccf}, the solid curve 
shows the CCF for cluster N5236-502 vs.\ the template star HR~4226.
The three dashed curves show the CCFs for the spectrum of another 
template star, HR~5645, broadened by 0 km/s (i.e.\ no broadening), 4 km/s 
and 8 km/s, vs.\ HR~4226. From this example, the cluster appears to have a 
velocity dispersion greater than 4 km/s, but less than 8 km/s. 
An important feature of the cross-correlation technique is that
the velocity dispersion of the science object is measured relative to
a template. Therefore, resolution effects and
any intrinsic broadening of the lines (e.g.\ due to macroturbulent
motions in the template star atmosphere) will cancel out, as long as
these effects are reproducible.  Also, because only the width of the 
CCF peak is used in the derivation of velocity dispersions, addition of 
a smooth continuum to the spectra (e.g.\ from early-type stars in the 
clusters) will not affect the measurements.

  The amplitudes of the cross-correlation peaks are listed in 
Table~\ref{tab:templates} for each template vs.\ the two clusters, 
cross-correlating over the wavelength range 7310\AA --7570\AA\ (UVES-RL(2), 
see below).  The template star spectra were broadened with a 
series of Gaussians with dispersions separated by steps of 1 km/s.  Except 
for the three G-type supergiants (see below), we used all combinations of 
the template stars in Table~\ref{tab:templates} to produce estimates of the 
velocity dispersions.  For each combination of template stars, the velocity 
dispersions of the cluster spectra were determined by interpolation between 
the dispersions of the two best-matching broadened template spectra.  
Including the three G-type stars in the cross-correlation analysis 
significantly increased the scatter, and since the cross-correlation signal 
also tends to be weaker than average for these stars we have excluded them
from the template sample. Presumably the poorer match by the G stars
indicates that the cluster spectra are indeed dominated by later-type stars.

  The mean and median values of the velocity dispersions derived for
all combinations of template stars are listed in Table~\ref{tab:vd},
along with the dispersions of the measurements, the mean peak values
of the CCFs ($h$) and the S/N of the cluster spectra.  The analysis
was carried out for several different wavelength ranges: 4520--4850 \AA\ 
(UVES-B), 6770--6850 \AA\ and 6940--7110 \AA\ (UVES-RL(1)), 7310--7570 \AA\ 
(UVES-RL(2)), 7710--8070 \AA\ (UVES-RL(3)) and 8680--8880\AA\ (UVES-RU).
In the UVES-RL(1,2) entry, the regions covered by UVES-RL(1) and UVES-RL(2) 
were both included at the same time in the cross-correlation analysis. 
The choice of these wavelength ranges was dictated partly by the need for 
adequate S/N (on the blue side) and partly by the requirement to avoid 
spectral regions that are strongly affected by night sky lines and
atmospheric absorption bands (on the red side). 
While the UVES blue spectra extend well shortwards of 4500\AA, the 
contribution of red supergiants to the integrated light drops off rapidly 
at shorter wavelengths.  We made some
crude estimates of the relative contributions of cool and hot stars at 
various wavelengths using Girardi et al.\ (\cite{gir2000}) isochrones.
The exact ratio is very sensitive to the age (and also quite
model-dependent), especially for 
N5236-805, but if we define $L_{\rm cool}/L_{\rm tot}$
as the relative fraction of the total luminosity contributed by
stars with $T_{\rm eff}<5000$ K, then we find that this ratio decreases
from $>50$\% in the $I$-band to $\sim25$\% in $V$, 5\%--10\% in 
the $B$-band, and $<1$\% in $U$ at an age of about $10^7$ years. Note, 
however, that the cross-correlation analysis does \emph{not} rely on any 
individual strong lines, but instead utilizes the numerous weaker lines 
present in the spectra of red supergiants. 

\subsection{Uncertainties}

  It is gratifying to see that the cross-correlation technique produces 
fairly consistent results in the different wavelength ranges, with a scatter 
of less than 10\% between the different measurements. {\rm Even the UVES-B
setting produces consistent results, although the cross-correlation
signal here is weak}.  Furthermore, the scatter 
of the individual measurements within one wavelength range is always less than 
1 km/s, suggesting that the choice of template star is not critical.  This is 
an important point, because the turbulent velocities in the atmospheres of 
bright red giants and supergiants are comparable ($\approx$ 10 km/s) to the 
velocity dispersions 
we are trying to measure here (Smith \& Dominy \cite{sd79}; Gray \& Toner 
\cite{gt87}).  The small scatter in the velocity dispersion measurements 
suggests that the intrinsic scatter in the macroturbulent velocities among the 
template stars must be relatively small as well ($<1$ km/s).  
Still, the macroturbulence 
is an important source of uncertainty to keep in mind, and it cannot be
excluded that the red supergiants in the clusters have systematically 
different properties in this respect than the local Galactic stars used as
templates.  If this is the case, our velocity dispersion measurements might 
be biased in either direction. 
For N5236-502, this concern is compounded
by the fact that the velocity dispersion for this cluster is uncomfortably
close to the intrinsic broadening of the lines in the supergiant stars.
For N5236-805 there is another problem, namely, that the low age makes it 
hard to find suitable template stars. Stellar isochrones from
Girardi et al.\ (\cite{gir2000}) predict that the red supergiants have
$M_V\sim-6$ at log(age)$\approx$7.1, much brighter than any of the templates
in Table~\ref{tab:templates}.  To test whether any strong dependence on
the template star luminosity class exists, we repeated the analysis for
the luminosity class Ib and II templates separately and found that the 
results agreed within 0.1--0.2 km/s, well within the random uncertainty
of the measurements. 
We note, however, that a similar analysis of Keck/HIRES spectroscopy
of four massive clusters studied by Larsen, Brodie \& Hunter (\cite{lbh04}) 
yielded a somewhat greater difference of $\sim0.6$ km/s between velocity 
dispersions derived from luminosity class Ia/Ib and II template stars,
with Ia/Ib templates yielding lower velocity dispersions.  That 
paper also includes a more detailed analysis of the dependence of the 
derived velocity dispersion on the template star properties, including 
spectral type and luminosity class.  If there is a general trend for the 
macroturbulent velocities to increase with luminosity then we could have 
overestimated the velocity dispersion (and, consequently, the
mass-to-light ratio) for N5236-805.  A more detailed 
investigation of this issue would clearly be of great value.

  As our final estimates of the line-of-sight velocity dispersions we adopted
a weighted average of the mean values in each wavelength range. We
calculated two weighted averages for each cluster, one using the r.m.s.
dispersion within each wavelength range for the weights, and another one
using the mean amplitude of the cross-correlation peak.  The result did 
not depend strongly on the choice of weighting scheme and we simply adopted
the mean of the two weighted averages.  For the error estimates, we 
adopted the larger of the estimated standard errors on the weighted averages. 
The final adopted line-of-sight velocity dispersions are
$5.5\pm0.2$ km/s for N5236-502 and $8.1\pm0.2$ km/s for N5236-805.

\section{Structural parameters}

\begin{table}
\caption{Structural parameters from ISHAPE fits to HST/WFPC2 images.
  Numbers in this table are for a fitting radius of $3\arcsec$
  (30 pixels). For N5236-502 and N5236-805 the 'V' entries refer
  to F606W and F547M data, while 'I' refers to F814W for both clusters.}
\label{tab:struct}
\begin{tabular}{lrr} \\ \hline
                       &  N5236-502   & N5236-805    \\ \hline
King models ($V$)        &              &              \\
\ FWHM                 &       -      & $0\farcs054$ \\ 
\ $r_t/r_c$            &       -      &  $75\pm20$   \\
\ \reff                &       -      & $0\farcs122\pm0.02$ \\
King models ($I$)        &              &              \\
\ FWHM                 & $0\farcs14$  & $0\farcs054$ \\ 
\ $r_t/r_c$            & $100\pm20$  &     $91\pm20$   \\
\ \reff                & $0\farcs36\pm0\farcs05$ & $0\farcs133\pm0\farcs02$ \\
EFF    models ($V$)      &              &              \\
\ FWHM                 &       -      & $0\farcs084$ \\ 
 \ Slope parameter      &       -      & 1.32     \\
 \ \reff (\rmax = $\infty$)
                        &              & $0\farcs140$ \\
 \ \reff (\rmax = $3\arcsec$)
                        &              & $0\farcs124$ \\
EFF    models ($I$)      &              &              \\
\ FWHM                 &  $0\farcs17$ & $0\farcs090$ \\ 
\ Slope parameter      &   1.14       & 1.28  \\
\ \reff (\rmax = $\infty$) 
                       & $1\farcs101$ & $0\farcs176$ \\
\ \reff (\rmax = $3\arcsec$) 
                       & $0\farcs339$ & $0\farcs144$ \\
Adopted                &              &               \\
\ \reff                & $0\farcs35\pm0\farcs05$ & $0\farcs13\pm0\farcs02$ \\
\hline
\end{tabular}
\end{table}

  Structural parameters for the clusters were measured on HST 
images obtained under programmes 5971 (for N5236-502) and 
8234 (for N5236-814). The data were downloaded from the archive at
STScI and processed on-the-fly by the standard pipeline.  The 
individual exposures were then combined with the CRREJ task in the
STSDAS package in IRAF.  For programme 
5971, four exposures in F814W ($2\times1000\mathrm{s} + 
2\times1300\mathrm{s})$ and two exposures in F606W (1100 s $+$ 1200 s) were 
available, but the central pixels of N5236-502 were saturated on all but 
the short (1000 s) F814W exposures, so these were the only exposures used for 
the 
analysis of structural parameters for this cluster.  The dataset for programme 
8234 consisted of 3 exposures in F547M ($180+350+400$ s) and 3 exposures 
in F814W ($160+200+350$ s), allowing us to carry out separate analyses in the 
two bands. Both clusters are imaged on the WF chips of their
respective HST datasets.

  Interestingly, the HST and VLT images reveal a smaller companion cluster, 
separated
by $1\farcs2$ from  N5236-502, which is only hinted at in the images from 
the Danish 1.54 m telescope as a slight elongation of the cluster image. 
The companion cluster is clearly resolved on the HST images.  The projected 
separation between the two clusters corresponds to only 26 pc, so they might 
constitute a real physical pair.  From photometry in an $r=5$ pixels aperture 
on the HST images we find a magnitude difference between the two clusters of 
$\Delta_\mathrm{F606W}$ = 1.34 mag and $\Delta_\mathrm{F814W}$ = 1.62 mag. 
Thus, the smaller cluster appears somewhat bluer, although the magnitude 
difference in F606W may be slightly underestimated due to the saturation 
of the brighter cluster. If the clusters are a physical pair, the colour
difference might be due to differences in the reddenings towards the
clusters, which is not too unlikely considering that their position 
coincides with a conspicuous dust lane (Fig.~\ref{fig:m83}).  Below we 
estimate an age for N5236-502 of about 100 Myr (Section~\ref{sec:phot}), 
so the clusters probably did not form out of material associated with
this dust lane.

  At the adopted distance of M83, the image scale of the WF chips
corresponds to 2.17 pc pixel$^{-1}$. This is comparable to the
typical half-light radii of stellar clusters, so a correction for the
WFPC2 point spread function (PSF) is necessary. We used the ISHAPE code 
(Larsen \cite{lar99}) to fit analytic King (\cite{king62}) models and 
``EFF'' (Elson et al.\ \cite{eff87}) profiles of the form
\begin{equation}
  P(r) \propto \left(1 + (r/r_c)^2\right)^{-\alpha}
  \label{eq:moffat}
\end{equation}
to the images.  ISHAPE determines the best fitting shape parameters by 
iteratively convolving the analytic model with the PSF, adjusting the FWHM 
and envelope slope parameter $\alpha$ (or concentration parameter 
$c$ = $r_t/r_c$ in the case of King models) until the best fit to the 
observed cluster images are obtained.  The PSF was modelled using version 
6.1 of the TinyTim code (Krist \& Hook \cite{kh97}) and the fits were
carried out using a fitting radius of $3\arcsec$ (30 pixels).  Integration of 
Eq.~(\ref{eq:moffat}) yields the following expression for the half-light 
or \emph{effective} radius, \reff, of the EFF models:
\begin{equation}
  \reff \, = \, r_c \, \sqrt{(1/2)^{\frac{1}{1-\alpha}} -1}
  \label{eq:reff1}
\end{equation}
for a profile extending to infinity and with $\alpha>1$, and
\begin{equation}
  \reff \, = \, r_c \left[\left\{\frac{1}{2}\left[
               \left(1+\frac{\rmax^2}{r_c^2}\right)^{1-\alpha}+1\right]
                 \right\}^{\frac{1}{1-\alpha}}-1\right]^{1/2}
  \label{eq:reff2}
\end{equation}
for a profile truncated at an outer radius \rmax .  For $\alpha$ values
significantly greater than 1, \reff\ does not depend strongly on the 
adopted \rmax, but this convenient behaviour breaks down as $\alpha$
approaches or even decreases below unity.


  The results of the profile fits are listed in Table~\ref{tab:struct}.
The ISHAPE code does not provide formal error estimates on the fitted 
parameters, and many of the uncertainties are likely to be systematic
rather than random. An estimate of the reproducibility of the results
can be obtained by repeating the fits several times with different
initial guesses for the fit parameters and for different fitting radii. Such 
an exercise shows that the FWHM generally reproduces within 0.1 pixel 
(0\farcs01) and the EFF 
$\alpha$ parameter is stable within a few percent.  The King profile 
concentration parameter is uncertain by at least 20\%.
In the case of N5236-502, the King and EFF fits agree 
reasonably well on the FWHM ($0\farcs14$ and $0\farcs17$), corresponding to 
3.0 and 3.7 pc. The King fits yield an effective radius
of $\reff=0\farcs36$ (7.8 pc) with a formal uncertainty of about $0\farcs05$ 
(1 pc),
mostly due to the error on the concentration parameter. For the EFF
fits, the effective radius is sensitive to the adopted outer radius of the 
cluster because the slope $\alpha$ is only slightly greater than 1. If we 
adopt the same \rmax\ (3\arcsec) used for the fit itself, 
the resulting $\reff = 0\farcs34$. For $\rmax=2\arcsec$ and $\rmax=5\arcsec$,
Eq.~(\ref{eq:reff2}) yields $\reff=0\farcs30$ and $\reff=0\farcs39$.
In the extreme case of a profile extending to infinity, Eq.~(\ref{eq:reff1})
yields an effective radius of $1\farcs10$.  For a smaller fitting radius 
of $2\arcsec$ the FWHM increases slightly (FWHM$\approx0\farcs18$), balanced 
by an increase in $\alpha$ ($\sim1.21$) and leading to a slight net 
decrease (by about $0\farcs03$ relative to the fits within $3\arcsec$) in 
\reff .  For N5236-805, the envelope
slope is steeper and \reff\ is therefore less model-dependent. The fits all 
agree on an effective radius close to $0\farcs13$. Thus, we adopt 
$\reff=0\farcs35$ (7.6 pc) and $0\farcs13$ (2.8 pc) for N5236-502 and 
N5236-805 with uncertainties of $0\farcs05$ (1.1 pc) and $0\farcs02$
(0.4 pc). Interestingly, 
the cluster near the centre of NGC~5236 is much more compact than the one 
located further out.


\section{Photometry and cluster ages}
\label{sec:phot}

\begin{table*}
\caption{Ground-based and HST photometry for the clusters. A 
  foreground reddening correction of $A_B=0.284$ mag has been applied.}
\label{tab:photometry}
\begin{tabular}{lccccccc} \\ \hline
%
Cluster   & R.A.     &   Decl.     &  $V$   & \ub    &  \bv  &  \vi  &  $V$(HST) \\
          & \multicolumn{2}{c}{2000.0} &    &        &       &       & $r=2\farcs0$ \\ \hline
N5236-502 & 13:37:06.12 & $-$29:53:18.4 & $17.51\pm0.01$ & $-0.11\pm0.03$ & $0.50\pm0.02$ & $0.89\pm0.02$ & $17.43\pm0.01$ \\
\multicolumn{3}{l}{N5236-502 (VLT/FORS)} &  $17.48\pm0.01$ & $-0.12\pm0.01$ & $0.39\pm0.01$ & $0.73\pm0.01$ & \\
N5236-805 & 13:37:01.82 & $-$29:52:13.1 & $16.60\pm0.02$ & $-0.55\pm0.02$ & $0.19\pm0.03$ & $0.65\pm0.04$ & $16.83\pm0.01$ \\ \hline
\end{tabular}
\end{table*}

\begin{table}
\caption{Age and reddening estimates for the clusters. For N5236-502 we
  used the VLT/FORS photometry. The $A_B$ values listed here are in
  addition to our adopted foreground reddening of $A_B=0.284$ mag.}
\label{tab:age_ab}
\begin{tabular}{lcccc} \\ \hline
     & \multicolumn{2}{c}{N5236-502} & \multicolumn{2}{c}{N5236-805} \\ 
             &  Log(age) &  $A_B$        & Log(age) & $A_B$       \\ \hline
S-sequence   &    7.99   &  0.71         &  7.42    & 0.24        \\
$Z=0.008$    &           &               &          &             \\
\ $UBV$      &    7.90   &  1.31         &  6.80    & 0.20        \\
\ $UBVI$     &    8.06   &  1.02         &  7.16    & 1.00        \\ 
$Z=0.02$     &           &               &          &             \\
\ $UBV$      &    8.01   &  1.13         &  7.10    & 0.56        \\ 
\ $UBVI$     &    8.11   &  0.86         &  6.82    & 1.07        \\
$Z=0.05$     &           &               &          &             \\
\ $UBV$      &    7.86   &  1.42         &  7.12    & 1.41        \\ 
\ $UBVI$     &    8.06   &  0.78         &  6.66    & 1.55        \\ 
Adopted      & $8.0\pm0.1$ & $1.0\pm0.2$ & $7.1\pm0.2$ & $1.0\pm0.5$ \\ \hline
\end{tabular}
\end{table}

\begin{figure}
\includegraphics[width=85mm]{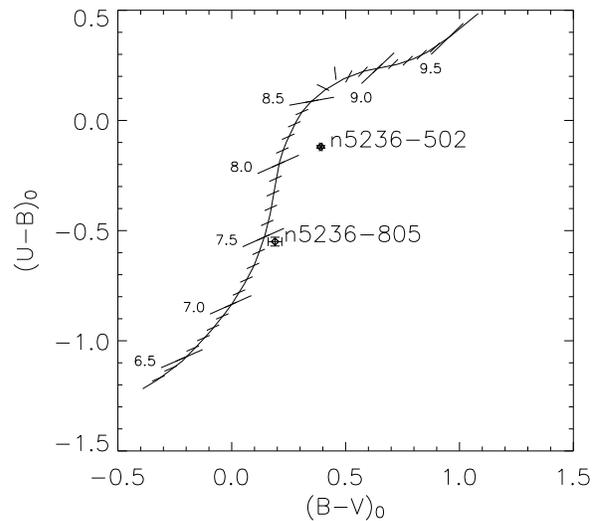}
\caption{\bv, \ub\ two-colour diagram indicating the location of
  the two clusters and the $S$-sequence of Girardi et al.\ (\cite{gir95}).
  The logarithm of the age is indicated along the sequence.
  The photometry has been corrected for foreground reddening only.
}
\label{fig:bv_ub}
\end{figure}

\begin{figure}
\includegraphics[width=85mm]{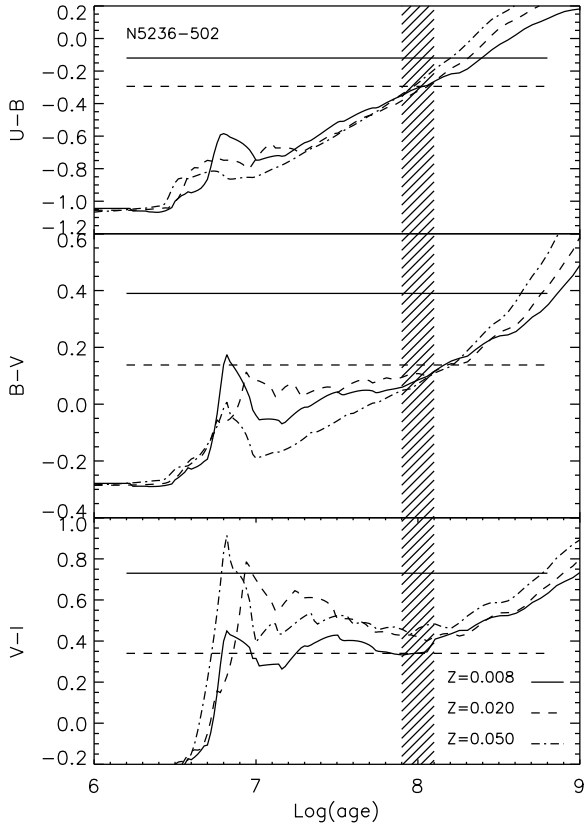}
\caption{Comparison of Bruzual \& Charlot model colours with observed
  colours for cluster N5236-502. The horizontal solid and dashed
  lines indicate the cluster colours corrected for foreground reddening 
  only and assuming an additional $A_B=1.0$ mag, respectively. The
  hatched areas indicate the ages and adopted uncertainties.
  The model colours are shown for $Z=0.008$, $Z=0.020$ and $Z=0.050$.
}
\label{fig:cfig502}
\end{figure}

\begin{figure}
\includegraphics[width=85mm]{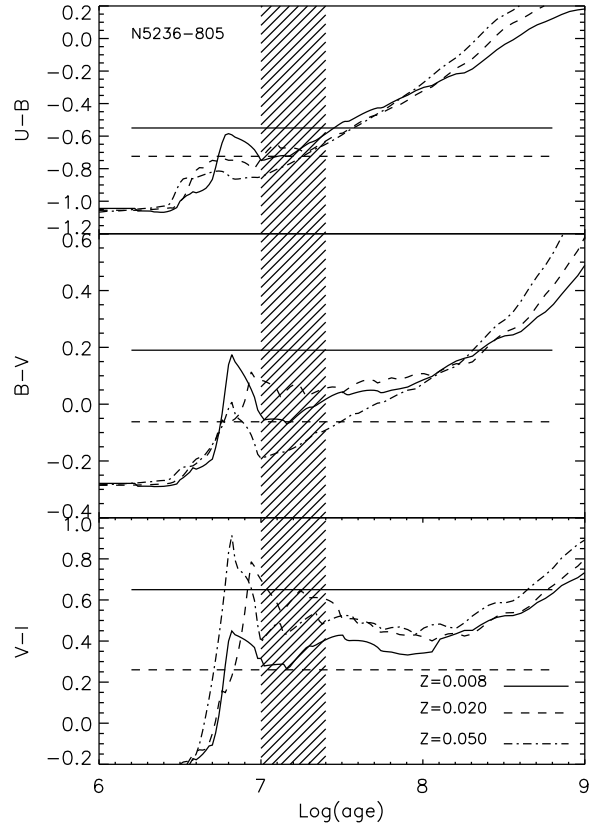}
\caption{Same as Fig.~\ref{fig:cfig502}, except that cluster N5236-805
 is shown.
}
\label{fig:cfig805}
\end{figure}


  Age estimates were obtained from $UBVI$ colours.  Table~\ref{tab:photometry} 
lists the integrated $V$ magnitudes and \ub, \bv\ and \vi\ colours for each 
cluster, corrected for a Galactic foreground extinction of $A_B = 0.284$ mag 
(Schlegel et al.\ \cite{sch98}, provided by the \emph{NASA/IPAC Extragalactic
Database} (NED)). For both clusters, we obtained
photometry from images taken with the Danish 1.54 meter telescope at ESO, 
La Silla, using an $r=4$ pixels ($1\farcs56$) aperture (see Larsen 
\cite{lar99} for details concerning the photometric calibration).  In addition 
to the data from the Danish telescope, we also list photometry for N5236-502 
from VLT FORS1/FORS2 commissioning data (Comer{\'o}n \cite{com01}). These 
data have the advantage of much better seeing ($\sim0\farcs6$ rather than 
$1\farcs5$), but lack an adequate photometric calibration. Therefore, the FORS 
data were tied to the photometric zero-points of the Danish 1.54 m data using 
a number of isolated stars outside the main body of NGC~5236.  The better 
seeing in the FORS images allows a more local measurement of the 
sky background and a smaller aperture radius (4 pixels = $0\farcs8$) for the 
cluster photometry, reducing possible systematic effects on the colours 
from the nearby companion cluster.  Unfortunately, N5236-805 is saturated in 
the FORS images, so for this cluster we only have the Danish 1.54 m data to 
rely on. Both the Danish 1.54 m dataset and the FORS1/FORS2 data include
H$\alpha$ imaging, from which it is clear that neither cluster has any
line emission associated with it.

  Table~\ref{tab:photometry} also includes estimates of the $V$ magnitudes
from the HST photometry, obtained in an $r=2\arcsec$ aperture and
using the zero-points in the \emph{WFPC2 instrument handbook}
(Biretta et al.\ \cite{bir96}). While
the central pixels are saturated for N5236-502, these contain only 
a small fraction of the total flux, and the fact that the 1000 s F814W 
exposures do not reach saturation furthermore suggests that the F606W 
exposures (of comparable length) are not strongly saturated. Overall,
the $V$ magnitudes from different sources agree within 0.1-0.2 mag,
which probably provides a better estimate of the true uncertainty due
to the background determination, aperture size etc, than the formal
errors due to photon statistics. It should be noted that our errors
on the HST photometry do not include uncertainties on the
photometric zero-points, which may be as large as $\sim0.05$ mag
according to the WFPC2 instrument handbook. For the remainder of 
this paper, we adopt the HST values for the total $V$ magnitudes of both 
clusters, the VLT FORS1/FORS2 colours for N5236-502 and the colours from 
the Danish 1.54 m data for N5236-805.

  In Figure~\ref{fig:bv_ub} we show the photometry for the two clusters in 
a (\bv, \ub) two-colour diagram, together with the ``S''-sequence defined 
by Girardi et al.\ (\cite{gir95}). The S-sequence is basically the mean locus 
of LMC clusters in the (\bv, \ub) plane, and defines an age sequence. The 
tick marks along the sequence indicate the optimal direction in which to shift 
clusters onto the sequence, considering reddening effects as well as stochastic 
colour variations. While the S-sequence has certain limitations, such as being 
defined only for LMC metallicity and restricted to the $\bv, \ub$ two-colour 
plane, some rough estimates of the ages and reddenings of the two clusters 
can be drawn already from a visual inspection of Fig.~\ref{fig:bv_ub}. The 
logarithm of the ages of N5236-502 and N5236-805 appear to be about 7.4 
(25 Myr) and 8.0 (100 Myr), respectively. For N5236-805, only a small shift 
is required to bring the cluster on the S-sequence, suggesting a small 
reddening correction, while a larger shift is necessary for N5236-502. The 
formal S-sequence age- and reddening estimates for the two clusters are listed 
in Table~\ref{tab:age_ab}.

  In order to utilize the full information contained in the $UBVI$
photometry and assess the possible role of metallicity differences, we must 
rely on model calculations for the evolution of broad-band colours as a 
function of age. In Figs.~\ref{fig:cfig502} and \ref{fig:cfig805} we compare 
Bruzual \& Charlot (2000; priv.\ comm.) model calculations for three different 
metallicities 
($Z=0.008$, $Z=0.02$ and $Z=0.05$) with the observed colours of the two 
clusters. In each figure, the horizontal solid lines indicate the observed 
\ub, \bv\ and \vi\ colours (corrected for \emph{foreground} reddening only), 
while the horizontal dashed lines indicate the 
effect of an additional internal reddening in NGC~5236 of $A_B=1$ mag. For 
each metallicity, Table~\ref{tab:age_ab} lists the best-fit ages and 
internal $A_B$ values obtained by 
minimizing the r.m.s. difference between the model and observed colours, 
using both the full set of $UBVI$, as well as just $UBV$. For N5236-502, 
the comparison with model colours yields ages that are quite similar to 
those obtained via the S-sequence method, albeit with somewhat higher
reddening estimates. In the case of N5236-805 there are 
more pronounced differences, and from Fig.~\ref{fig:cfig805} it is clear that
the models are very metallicity-sensitive in the age range around
the best-fit value ($\sim10^7$ years) for this cluster. In particular, the 
``red loop'' in the broad-band colours around $10^7$ years is smoothed over 
in the S-sequence.  Thus, the age- and reddening estimates are more
uncertain for this cluster, although the absence of line emission associated
with the cluster argues against an age of much less than $\sim10^7$ years.

  Based on the various age estimates in Table~\ref{tab:age_ab}, we adopt 
estimates of log(age) = $8.0\pm0.1$ and $7.1\pm0.2$ and reddenings of 
$A_B=1.0\pm0.2$ and $A_B=1.0\pm0.5$ for N5236-502 and N5236-805. These 
age estimates are indicated by the shaded areas in Figs.~\ref{fig:cfig502} 
and \ref{fig:cfig805}.

\section{Modelling mass-to-light ratios}

\begin{figure*}
\includegraphics[width=160mm]{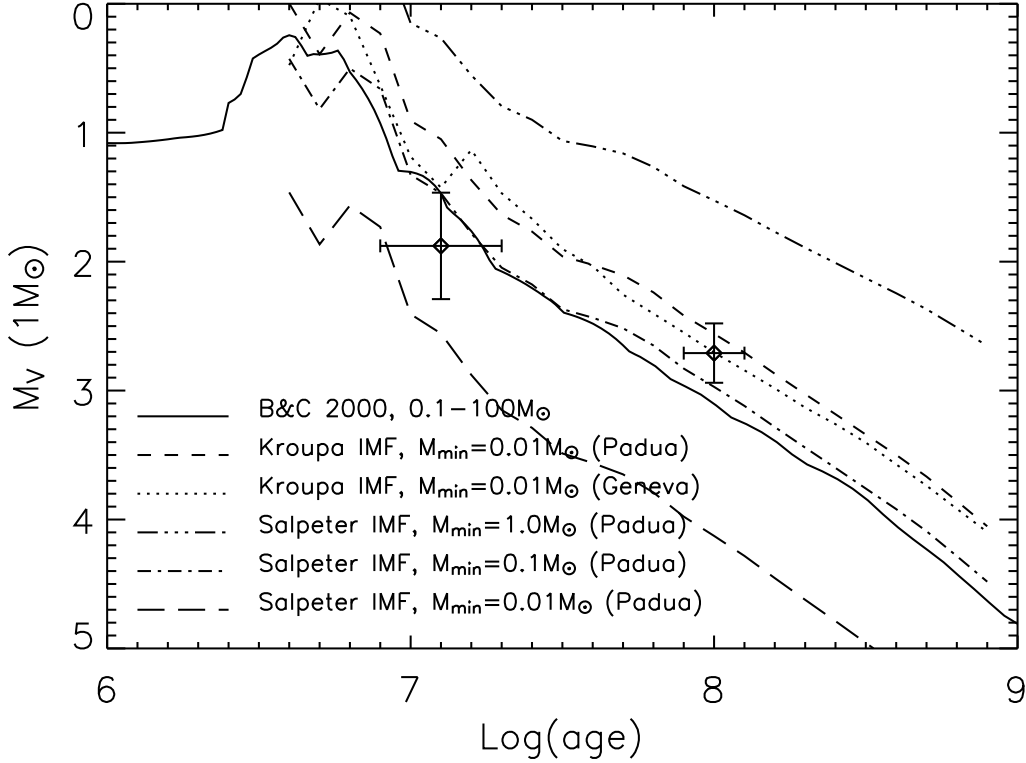}
\caption{Absolute $M_V$ magnitudes per solar mass computed for various
  combinations of stellar mass functions and isochrones. The solid line
  is from SSP models by Bruzual \& Charlot (2000), computed for a 
  Salpeter MF from 0.1--100$\msun$. The long-dashed, dotted-dashed and
  triple-dotted-dashed lines are for Salpeter MFs with lower mass limits 
  of $M_\mathrm{min}=0.01\msun, 0.1\msun$ and $1.0\msun$, computed from Padua 
  isochrones, while the short-dashed line is for a Kroupa MF 
  with $M_\mathrm{min}=0.01 \msun$. The dotted line is again for
  the Kroupa MF, but using isochrones from the Geneva group. All models 
  are for solar metallicity. Data for the two clusters are shown with
  error bars.
}
\label{fig:pmtol}
\end{figure*}

  The Bruzual \& Charlot models tabulate mass-to-light ratios in different 
passbands for various stellar MFs. However, in order to have the freedom to 
explore the effects of varying the MF and compare results for different 
stellar models, we calculated our own M/L ratios using isochrones from both 
the Padua group (Girardi et al.\ \cite{gir2000} and references therein) 
and the Geneva group (Lejeune \& Schaerer \cite{ls01}). In 
Fig.~\ref{fig:pmtol} we compare various calculations of the $M_V$ magnitude 
per solar mass for these simple stellar population models. As a reference, the 
solid line shows the Bruzual \& Charlot SSP models (which are based on Padua 
stellar models) for a Salpeter (\cite{salp55}) MF in the mass range 
0.1--100$\msun$ and solar metallicity. The dotted-dashed line shows our 
calculations for a Salpeter MF with a minimum mass $M_\mathrm{min}=0.1\msun$, 
also based on Padua isochrones. The overall agreement between our calculations 
and the Bruzual \& Charlot models is quite satisfactory and certainly adequate 
for our purpose. The minor differences may well be due to different treatment 
of stellar remnants (we do not count any mass above the endpoint of stellar 
evolution at a given age).  The short-dashed and dotted lines both show 
results for 
a Kroupa (\cite{kroupa02}) MF with $M_\mathrm{min}=0.01\msun$, but for Padua 
and Geneva isochrones, respectively. The triple-dotted-dashed and
long-dashed lines represent Salpeter MFs (based on Padua models) with
lower cut-offs at 1\msun\ and $0.01\msun$.  For the youngest ages 
(log(age)$<$7.2) there are substantial differences between the Padua and 
Geneva models, but otherwise the main difference between the various curves 
lies in the MF choice.

\section{Results}

\begin{table*}
\caption{Derived physical parameters. Uncertainties on $M_V$ are
  mainly due to the uncertainties on $A_B$. Photometric mass
  estimates are given for a Kroupa MF and for Salpeter MF
  with lower mass limit of $0.1\msun$. Central $V$-band surface 
  brightnesses ($\mu_{0,V}$) in mag arcsec$^{-2}$ and estimated core 
  densities ($\rho_0$) in \msun\ pc$^{-3}$ are also listed.}
\label{tab:phys}
\begin{tabular}{lrr} \\ \hline
                  & N5236-502    &  N5236-805   \\ \hline
$M_V$         &  $-11.57\pm0.15$ & $-12.17\pm0.37$   \\
Log (age)         &  $8.0\pm0.1$ &  $7.1\pm0.2$ \\
$M_\mathrm{phot}$ (Kroupa) [\msun]& $(4.49\pm 0.86)\times10^5$ & 
                    $(1.93\pm 1.42)\times10^5$ \\
$M_\mathrm{phot}$ (Salp) [\msun]
                  & $(6.56\pm 1.26)\times10^5$ & 
                    $(2.84\pm 2.06)\times10^5$ \\
\reff\ [pc]       &  $7.6\pm1.1$ & $2.8\pm0.4$ \\
$\mu_{0,V}$ [mag arcsec$^{-2}$] & $14.1\pm0.2$ & $12.0\pm0.5$ \\
$\rho_0$ [\msun\ pc$^{-3}$] & $(2.8\pm1.0)\times10^3$ & $(1.6\pm1.1)\times10^4$ \\
$M_\mathrm{vir}$ [\msun] & $(5.15\pm 0.83)\times10^5$ &
                    $(4.16\pm 0.67)\times10^5$ \\ 
$M_\mathrm{phot}/M_\mathrm{vir}$ (Kroupa) & $0.87\pm0.22$ & $0.46\pm0.34$  \\
$M_\mathrm{phot}/M_\mathrm{vir}$ (Salp) & $1.27\pm0.32$ & $0.68\pm0.51$  \\
\hline
\end{tabular}
\end{table*}

  From application of the virial theorem, the velocity dispersion
$v_m$, total mass $M$ and half-mass radius $r_h$ of a star cluster
are related as
\begin{equation}
  v_m^2 \approx \frac{0.4 G M}{r_h}
  \label{eq:vir1}
\end{equation}
(Spitzer \cite{spit87}, p.\ 11--12).  The 3-dimensional half-mass radius 
$r_h$ is related to the observed half-\emph{light} radius \reff\ roughly 
as $r_h = 1.3 \reff$, and with the line-of-sight velocity dispersion
given as $v_x^2 = \frac{1}{3}v_m^2$ we can express Eq.~(\ref{eq:vir1}) in 
terms of observable quantities:
\begin{equation}
  M_\mathrm{vir} \, \approx \, 9.75 \frac{\reff \, v_x^2}{G}.
  \label{eq:vir2}
\end{equation}
  The exact value of the constant in front of Eq.~(\ref{eq:vir1}) depends on 
the detailed radial profile of the cluster, but the value 0.4 should
be accurate to $\sim$10\% or better for most relevant cases 
(Spitzer \cite{spit87}). We denote the mass derived from 
Eq.~(\ref{eq:vir2}) $M_\mathrm{vir}$ to distinguish it from a
``photometric'' mass estimate ($M_\mathrm{phot}$), based on SSP models.
The above relations are valid for spherically symmetric clusters with an 
isotropic velocity distribution, which are in virial equilibrium.
Another implicit assumption is that the cluster light is a good tracer
of the underlying mass distribution.

  With these above caveats in mind, Table~\ref{tab:phys} compares the 
virial mass estimates (from Eq.~\ref{eq:vir2}) with photometric
masses for the two clusters.  The photometric mass estimates
are based on Padua isochrones and are calculated both for
Kroupa and Salpeter ($M_{\rm min}=0.1\msun$) MFs. The uncertainties on the
photometric mass estimates are based on the estimated uncertainties on 
the age (affecting the M/L ratio) and on the reddening (affecting 
$M_V$). These uncertainties are treated as uncorrelated although
this is, strictly speaking, not true.  Errors on the virial masses
are based on the error estimates for the velocity dispersion in
Table~\ref{tab:vd} and on the sizes. The uncertainty on the distance
to NGC~5236 has not been included in the error budget, but the
comparison of photometric and virial mass estimates is not very sensitive
to the exact distance (M/L ratios scale linearly with distance).
The estimated $\pm0.15$ mag error on the distance modulus quoted by 
Thim et al.\ (\cite{thim03}) corresponds to an uncertainty of about 
7\% on the ratio of $M_\mathrm{phot}/M_\mathrm{vir}$. For reference,
we also list the central $V$-band surface brightness ($\mu_{0,V}$) and
the estimated core density $(\rho_0)$ for each cluster, using the
analytic expressions relating these quantities to the mass-to-light ratios
and sizes, luminosities and slope parameters of the EFF profiles given 
in Larsen, Brodie \& Hunter (\cite{lbh04}).

  For N5236-502, the agreement between photometric and virial mass estimates 
is remarkably good. The virial mass estimate falls between the Kroupa and 
Salpeter photometric mass estimates, but is compatible with both within the 
$1\sigma$ errors. In Fig.~\ref{fig:pmtol}, the cluster actually falls right 
on top of the Kroupa curve calculated for Geneva isochrones.  An MF with 
a significant deficiency of low-mass stars, such as a Salpeter MF
truncated at $1\msun$, can be ruled out.
For N5236-805 the difference between the Kroupa and virial masses is
somewhat larger than for N5236-502. The virial mass is, in fact, 
somewhat \emph{greater} than expected for a Kroupa MF, implying an
excess of low-mass stars if the numbers are taken at face value.
The photometric mass estimates are less secure for this cluster, however, 
mainly due to the uncertainties in the stellar models around this age, and 
the inherent uncertainties in the age- and reddening determination due to 
the irregular behaviour of the colours and the strong metallicity dependence, 
even if the models were perfect.  Again, a Salpeter MF with a cut-off at
1$\msun$ can be safely ruled out. The assumption of virial equilibrium
may also be less secure for this cluster, due to its young age. We may
also have underestimated our errors on the velocity dispersion, cluster
size, or both.  In conclusion, there is probably no significant discrepancy 
between the virial mass estimate for N5236-805 and the assumption that the 
mass distribution of the cluster member stars follows a Kroupa MF.

  How reliable are the velocity dispersions derived from integrated
spectroscopy of the clusters?  One concern is that the velocity dispersions
might be affected by small number statistics, i.e.\ the integrated
light might be dominated by a few luminous supergiants. To address
this question, we carried out Monte-Carlo simulations of the
colour-magnitude diagrams of the two clusters, drawing stellar masses
at random from a Kroupa MF until the total masses equaled the virial
mass estimates in Table~\ref{tab:phys}. For each ``star'' in these
synthetic clusters, we then looked up the corresponding colour and
luminosity in the Padua isochrones. For N5236-502 and N5236-805,
we estimate a total of 550 and about 100 red supergiants, respectively.
The smaller number of red supergiants in N5236-805 is mostly due to its 
younger age and correspondingly higher main-sequence turn-off mass (about 
13 \msun\ vs.\ 5 \msun ). Thus, we do not expect the velocity dispersions 
or other integrated properties of the clusters to be severely affected 
by small number statistics.



\section{Conclusions}

  We have measured velocity dispersions, structural parameters and broad-band 
colours for two stellar clusters in the nearby spiral NGC~5236 (M83). Using
these data, we have compared photometric mass estimates, based on Padua 
stellar isochrones and various assumptions about the stellar mass 
function (MF), with virial mass estimates for the two clusters.  We conclude 
that the observed mass-to-light ratios are consistent with a Kroupa-type MF, 
and we can rule out an MF with a significant deficiency of low-mass stars 
relative to the Kroupa MF.  We derive virial mass estimates of 
$(5.2\pm0.8)\times10^5\msun$ (N5236-502) and 
$(4.2\pm0.7)\times10^5\msun$ (N5236-805), comparable 
to or even somewhat higher than the typical masses of old \emph{globular} 
clusters. The velocity dispersions of these clusters (5.5 km/s and
8.1 km/s) are probably close to the limit at which it becomes impractical 
to measure them from integrated light. For even lower velocity dispersions,
systematic uncertainties due to variations in the turbulent velocities of
the red supergiant stars, mass segregation, and other factors, will 
pose a formidable challenge.

While one of the clusters is located near the nucleus of NGC~5236 
and may disrupt on relatively short timescales, the other is located in the 
disk of the galaxy at a projected galactocentric distance of $1\farcm8$, 
corresponding to 2.3 kpc, and has an estimated age of about 100 Myrs.  The
masses of these clusters are comparable to those of the luminous young
clusters (sometimes referred to as ``super star clusters'') encountered
in merger galaxies and starbursts, but they are much higher than that
of any young open cluster observed in the Milky Way today. These clusters
provide a striking example of the fact that such clusters can still 
form in the disks of some apparently normal, undisturbed spiral galaxies.

\begin{acknowledgements}
  We thank F.\ Comer{\'o}n for supplying us with his reduced FORS1/FORS2 
  imaging data, and the anonymous referee for a number of useful comments.
  This research has made use of the NASA/IPAC Extragalactic 
  Database (NED) which is operated by the Jet Propulsion Laboratory, 
  California Institute of Technology, under contract with the National 
  Aeronautics and Space Administration. T.\ R.\ gratefully acknowledges
  support from the Chilean Center for Astrophysics FONDAP No.\ 15010003.
\end{acknowledgements}

\end{document}